\documentstyle[preprint,aps,prl,floats,amssymb]{revtex}

\def\mc{{\cal M}_{\rm cl}}

\def\tr{{\rm tr\, }}
\def\c{{\Bbb C}}
\def\A{{\cal A}}
\def\a{\alpha}
\def\b{\beta}
\def\g{\gamma} 
\def\Q{\tilde{Q}}

\def\dd{\hbox{\kern0.3em/\kern-0.7em /\kern0.5em}}
\def\lie#1{{\rm Lie}\left( #1 \right)}

\begin{document}
\tightenlines

\preprint{\vbox{
\hbox{UCSD/PTH 97--29}
\hbox{hep-th/9709161} }}

\title{'t~Hooft  Conditions in Supersymmetric
Dual  Theories}
\author{Gustavo Dotti}
\address{Department of Physics, University of California at San Diego,\\
9500 Gilman Drive, La Jolla, CA 92093-0319}
\date{September 1997}
\maketitle

\begin{abstract}
The matching of global anomalies of a supersymmetric 
gauge theory and its dual is seen to follow from similarities 
in  their classical chiral rings. These similarities 
provide a formula for  the dimension of the dual gauge group. 
As examples we derive 't~Hooft consistency conditions for the 
duals of supersymmetric QCD and $SU(N)$ theories with 
matter in the adjoint, and obtain the dimension of the 
dual groups.
\end{abstract}
\pacs{PACS}

One important constraint on the moduli space of vacua of supersymmetric gauge
theories~\cite{susy} is that the massless fermions in the low energy theory
should have the same flavor anomalies as the fundamental fields, i.e.
the 't~Hooft consistency conditions should be satisfied~\cite{thooft}. 
These conditions are  used as a test on the spectrum of massless fermions,  
usually obtained from  symmetry arguments 
and renormalization group flows. Two types  of theories 
have been  found:  for type I theories the classical 
moduli space $\mc$ or a suitable quantum modified version of it satisfies 
't~Hooft consistency conditions at every point; type II theories 
fail to satisfy these conditions at some points of $\mc$, and their 
quantum moduli space cannot just be a quantum modification of $\mc$.
It is believed that the IR sector of these theories at those points 
corresponds to a dual theory~\cite{s}.
The dual theory has a different gauge group and matter 
content, but the same flavor symmetry group.
As an example, consider supersymmetric QCD with $N_F$ flavors. 
When $N_F < N$ there is no supersymmetric vacuum in the quantum 
theory~\cite{ads}. When $N_F \geq N$, $\mc$ is described by mesons and 
baryons, which are gauge invariant polynomials in the microcopic 
fields, subject to some algebraic constraints. On general grounds 
it is shown that only the $N_F = N$ theories admit quantum deformations 
of the classical constraints. In fact, both $\mc$ in $N_F = N+1$ 
theories and a quantum deformation 
of $\mc$ in the $N_F = N$ case   
describe correctly the IR sector of these theories, 
these are type I theories~\cite{susy}. However, 
the origin of $\mc$ in $N_F \geq N+2$ theories fails 
't~Hooft's consistency test and is believed to correspond 
to  a  dual theory~\cite{s}.
Recently \cite{dm,dm2}, a mechanism responsible for 
flavor anomaly matching in type I theories was found 
and used to predict when a theory belongs to this group, 
avoiding explicit calculation of anomalies. Anomaly matching in 
s-confining theories \cite{sconf} (such as QCD with $N_F = N + 1$) 
and those obtained from them by 
integrating out matter fields (such as QCD with $N_F = N$), which have a 
quantum modified moduli space,  
follows from the  results in~\cite{dm}. In this letter we 
explore type II theories. 
The duality hypothesis in type II theories is supported by a number 
of consistency checks, of which the matching 
of global anomalies between both theories is believed 
to be a particularly stringent one. 
We will show that this matching follows 
from a sequence of applications of the results 
in~\cite{dm,dm2} and the relation between the classical moduli space 
of  both theories, intimately related to their classical chiral 
rings~\cite{dm2}.\\

We first review  the notation and state the results we need 
from 
\cite{dm}, a complete proof of them, together with a discussion 
on the classical moduli space from an algebraic geometry perspective 
can be found in~\cite{dm2}: 
$\phi^i, i=1,...,d_U$ 
is a point in the vector space $U$ of constant chiral field configurations 
of the UV theory, 
$G$ is the complexification of the gauge group $G_r$ of 
the theory, and $G \phi \subseteq U$ is the $G$ orbit 
of $\phi \in U$. 
If a tree level invariant superpotential 
$W(\phi)$ is added to the theory then $U^W \subseteq U$ 
denotes the set of critical 
points $dW = 0$, which contains complete $G$ orbits, as $W$ is $G_r$
invariant and holomorphic, therefore $G$ invariant.
 $V$ is the vector space spanned by a basic set of 
gauge invariant polynomials $\hat \phi ^i (\phi)$ in $U^W$ constructed 
out of the fundamental fields $\phi^i$. 
The tangent vector space of $U^W$ at the point $\phi_0$ 
is denoted $T_{\phi_0}U^W$. Under the natural isomorphism 
$T_{\phi_0}U \cong U$ we can regard $T_{\phi_0}U^W \subseteq U$ 
and expand a tangent vector in coordinates $\delta \phi^i$. 
 There is a natural 
map $\pi: U^W \to V$ by $\phi \to \hat \phi (\phi)$. 
The image 
$\mc = \pi(U^W)$ is called classical moduli space. The reason is 
that there is a unique closed orbit in every fiber $\pi^{-1}(\hat \phi), 
\hat \phi \in \mc$~\cite{dm}, and closed orbits are precisely 
those that contain a $D-flat$ point~\cite{lt}, therefore points in 
$\mc$ are in one to one correspondence with D-flat points 
satisfying $dW=0$, i.e, classical supersymmetric 
vacua. The differential $\pi'_{\phi_0}: T_{\phi_0}U^W 
\to T_{\hat \phi _0} \mc$ of $\pi$ at $\phi_0 \in U^W$ 
provides a linear map from the tangent of $U^W$ 
at $\phi_0$ to that of $\mc$ at $\hat \phi _0 = \pi (\phi_0)$. 
When $W=0$, $\mc$ is just the 
algebraic subset of $V$ defined by the constraints among the 
$\hat \phi ^i$~\cite{dm2}. As examples we introduce the two theories 
studied in this work:\\
{\em Supersymmetric QCD:} 
 the gauge group is $G_r = SU(N)$, its complexification 
$G = SL(N,\c)$. The matter fields 
 $\phi$ are the  quarks $Q^{i \alpha}, i=1,...,N_F$ in the 
fundamental of $SU(N)$ and the  antiquarks 
$\tilde{Q}_{\alpha j}, j=1,...,N_F$ in the dual of $SU(N)$, so the 
dimension of $U$ is $d_U = 2 N N_F$. A basic set of gauge 
invariant fields $\hat \phi^i$ is  
\begin{eqnarray} \label{M}
M^i_j &=& Q^{i \a} \Q_{\a j}  \\ \label{B}
B_{k_1 \cdots k_{N_D}}  &=&
    Q^{i_1 \a_1} Q^{i_2 \a_2} \cdots Q^{i_N \a_N} \epsilon_{\a_1 \a_2
\cdots \a_N} \epsilon_{i_1 i_2 \cdots i_N k_1 \cdots k_{N_D}}/N! \\
\label{Bt} \tilde{B}^{l_1 \cdots l_{N_D}} &=& \Q_{\a_1 j_1} 
\Q_{\a_2 j_2} \cdots
\Q_{\a_N j_N} \epsilon^{\a_1 \a_2 \cdots \a_N} \epsilon^{j_1 j_2 \cdots j_N
l_1  \cdots l_{N_D}}/N!,
\end{eqnarray}
where $N_D = N_F - N$; 
they span the vector space $V$ of dimension $N_F^2 + 2 N_F!/(N_D! N!)$.
$W=0$, then $\mc$ is the subset of $V$ defined by 
the algebraic constraints among the fields (\ref{M}-\ref{Bt}).\\
{\em $SU(N)$ theories with matter in the adjoint:} adding to the 
above theory a field $X$ in the adjoint (of $G = SL(N,\c)$) 
and a tree level superpotential $W = p(\tr X),$ 
$p$ a polynomial, we obtain the theories studied in~\cite{adjoint}. 
We will concentrate on the case $W = \tr X^3 /3$. 
Computations are  simpler if we drop the constraint 
tr $X = 0$, add a gauge singlet $T$ to the theory and replace 
the superpotential $W = $ tr $X^3/3$ with 
\begin{equation}
W = \frac{1}{3} \tr X^3 - \frac{T}{N} \tr X. 
\end{equation}
$T$ plays the role of a Lagrange multiplier. The equations 
$dW = 0$ defining $U^W$ are $\tr X = 0$ and $N X^2 = T {\Bbb I}$. 
Note that, although $T$ and $\tr X^k$ are independent gauge 
invariants on $U$, only $T$ is independent on $U^W$, 
where tr $X^{2s+1} = 0$ and tr $X^{2s} = T^s/N^{s-1}$, then  a 
basic set of gauge invariants is~\cite{adjoint}
\begin{eqnarray} 
M^i_j &=& Q^{i \a} \Q_{\a j} \label{Ma}\\
\label{N} 
N^i_j &=& Q^{i \a} X^{\b}_{\a} \tilde{Q} _{\b j}, \\ 
T (&=& X^{\a}_{\b} X^{\b}_{\a}),\label{T} \\
B^{i_1,...,i_{n_1};j_1,...,j_{n_2}} &=& 
Q^{i_1 \a_1} Q^{i_2 \a_2} \cdots Q^{i_{n_1} \a_{n_1}} 
X^{\b_1}_{\g_1} \cdots X^{\b_{n_2}}_{\g_{n_2}} Q^{j_1 \g_1} 
\cdots Q^{j_{n_2} \g_{n_2}} \epsilon_{\a_1 \cdots \a_{n_1} 
\b_1 \cdots \b_{n_2}}, \label{BB} \\ \label{BBt} 
\tilde{B}_{i_1,...,i_{n_1};j_1,...,j_{n_2}} &=& 
\tilde{Q}_{\a_1 i_1} \tilde{Q}_{\a_2 i_2} \cdots \tilde{Q}_{\a_{n_1} i_{n_1}} 
X_{\b_1}^{\g_1} \cdots X_{\b_{n_2}}^{\g_{n_2}} \tilde{Q}_{\g_1 j_1} 
\cdots \tilde{Q}_{\g_{n_2} j_{n_2}} \epsilon^{\a_1 \cdots \a_{n_1} 
\b_1 \cdots \b_{n_2}}, 
\end{eqnarray}
where $n_1 + n_2 = N$. These fields span the vector space $V$. \\

We now list the  results we need from~\cite{dm}, slightly generalized 
to the case $W \neq 0$, a detailed proof  
of these, together with a detailed description of $\mc$ from 
an algebraic geometry approach can be found in~\cite{dm2}.\\
{\bf Theorem I:}\ \ (i) Assume $G$ is totally broken at $\phi_0$ and 
the orbit $G \phi_0$ is closed (equivalently, $G \phi_0$ 
contains a D-flat point~\cite{lt,dm2}), then 
$\pi'_{\phi_0}: T_{\phi_0}U^W \to T_{\pi(\phi_0)}\mc$ is onto.\\
(ii) If also  $U^W$ is irreducible
 \footnote{This means that if the restriction 
to $U^W$ of the product of two polynomials in $U$ is 
zero, then the restriction of one of them must also 
be zero. The dimension of an irreducible algebraic set $X$ 
equals $\text{min}_{x \in X}$ dim $ T_x X$. The points of $X$ 
at which the tangent space has minimum dimension are said to be 
smooth~\cite{dm2,gw}.} and  $\phi_0 \in U^W$ is smooth, then 
$\mc$ is irreducible, $\hat \phi _0 = \pi(\phi_0)$ 
is smooth, and $\ker \pi'_{\phi_0} = \lie G \phi_0$. 
In particular, $T_{\hat \phi _0} \mc = \pi'_{\phi_0}(T_{\phi_0}U^W)$
and dim $\mc =$ rank $\pi'_{\phi_0}=$ 
dim $U^W -$ 
dim $G$.\\
{\bf Theorem II:}\ \ Let $\mc$ be the classical moduli space of a
supersymmetric gauge theory with gauge group $G$, 
superpotential $W$ and flavor symmetry $F$. It is
assumed that the gauge theory has no gauge or gravitational  anomalies, and the
flavor symmetries have no gauge anomalies. Let $\hat \phi_0 \in \mc$  be a
point in the classical moduli space. Assume there is a 
point $\phi_0 \in U^W$ in
the fiber $\pi^{-1}(\pi(\phi_0))$ of $\hat \phi_0$ such that\\
{\bf (a)} $G$ is completely broken at $\phi_0$.\\
{\bf (b)} $\pi'_{\phi_0}: T_{\phi_0}U^W \to T_{\hat \phi _0} \mc$ is onto.\\
{\bf (c)} $\ker \pi^\prime_{\phi_0} = \lie G \phi_0$. \\
If a subgroup $F_0 \subseteq F$ is unbroken at $\hat \phi_0$, then the 't~Hooft
consistency conditions for the $F_0^3$ 
flavor anomalies and the $F_0$ gravitational
anomalies are satisfied, i.e, the anomalies computed in the $F_0$  
invariant subspace $T_{\hat \phi _0} \mc$ match the corresponding 
anomalies in $U$.\\
{\bf Corollary:} If $\phi_0$ is a smooth point of the irreducible 
set $U^W$ that totally breaks $G$ and is D-flat (or has a closed 
$G$-orbit), then the anomalies of the  flavor subgroup 
unbroken at $\hat \phi _0 = \pi(\phi_0)$ match between $U$ 
and $T_{\hat \phi_0} \mc$. \\

The matching of flavor anomalies of a theory and its dual  
can be explained as follows: the dual theory has complexified
gauge group $G_D$, chiral configuration space space $U_D$, 
superpotential $W_D$ with critical points $U_D^{W_D}$. The 
$G_D$ invariant independent generators $(\hat \phi ^D)^i (\phi^D)$ 
in $U_D^W$ span a vector space   
$V_D$. The global symmetry 
group of the dual theory is the same as the flavor group 
$F$ of the original theory. There 
is an isomorphism $I: V_D \to V$ (the span 
of $G$ invariant generators of the original theory) which 
commutes with the action of $F$, therefore both $\mc$ and 
$\mc^D$ can be thought embedded in the same vector space.
In general, $\mc$ and $\mc^D$ are different. However, 
their intersection is nonempty and it is easy to find smooth points 
$\phi_i \in  U^W, \; 
\phi^D_i \in  U_D^{W_D}, i=1,...,s$ 
satisfying the 
hypothesis of theorem I, therefore those of theorem II, 
such that $\pi(\phi_i) = \hat \phi _i = \pi_D(\phi^D_i)$. 
We can calculate the tangent spaces to $\mc$ and $\mc^D$ at 
$\hat \phi _i$ using theorem~I, the superpotential 
$W_D$ is seen to be carefully chosen to satisfy 
$T_{\hat \phi _i} \mc = \pi'_{\phi_i}(T_{\phi_i}U^W) 
= (\pi_D)'_{\phi^D_i}(T_{\phi^D_i}U_D^{W_D}) = T_{\hat \phi _i} \mc ^D$. 
In some restricted cases we may have $\mc ^D = \mc$, 
then the above conditions are trivial. 
Denote $\A_F(X)$ the $F$ anomaly in the vector space $X$, 
and by $F_i$ the isotropy group of $\hat \phi _i$, 
i.e, the unbroken piece of $F$ at $\hat \phi _i$. 
Applying theorem  II twice gives: 
$$ \A _{F_i}(U) = \A _{F_i} (T_{\hat \phi_i} \mc) = 
\A _{F_i}(T_{\hat \phi_i} \mc ^D) = \A _{F_i}(U_D).$$
By choosing the points $\hat \phi_i$ appropriately, 
the matching of $U$ and $U_D$ anomalies for the full flavor 
group $F$ is implied by the matching of $F_i$ anomalies~\cite{dm2}.
We remark  that the argument does not require the classical moduli 
spaces of both theories to be the same,  but only to share 
the points $\hat \phi_i$, with the same tangent space at those points,  
and that these tangent spaces can be determined 
using theorem I, without even knowing the constraints 
that define $\mc$. \\

\noindent
{\bf Anomaly matching in dual SQCD theories:} 
The origin of $\mc$ in $N_F -2 \geq 
N \geq 2$ QCD theories displays the fully unbroken $SU(N_F)_L \times 
SU(N_F)_R \times U(1)_B \times R$ flavor symmetry; it is believed 
to correspond to  a dual theory with $N_F$ 
flavors of quarks in the fundamental of the gauge 
group $SU(N_D), N_D  = N_F - N$, 
$N_F$ flavors of antiquarks in the dual of 
the gauge group and additional gauge singlets $M^i_j$~\cite{s}. Note that 
$N_F - 2 \geq N_D \geq 2$ also. The  transformation 
properties of the fields in both theories are  summarized 
in the table below:

\begin{center}
\begin{tabular}{|c|c|c|c|c|c|}
\hline
 & $G_{(D)}$ & $SU(N_F)_L$ & $SU(N_F)_R$ & $U(1)_B$ & $R$ \\
\hline
$Q^{i\alpha}$ &$N$ &$N_F$ & $-$ & $1$ & $\frac{N_F-N}{N_F}$ \\
$\tilde Q_{j \alpha}$ & $\overline{N}$ & $-$ & $\overline{N}_F$ & $-1$ &
$\frac{N_F-N}{N_F}$ \\ \hline 
$q^{\a}_i$ & $N_D$ & $\overline{N}_F$ & $-$ & $\frac{N}{N_F-N}$ & 
$\frac{N}{N_F}$\\
$\tilde{q}^j_{\b}$ & $\overline{N}_D$ & $-$ & $N_F$ & $-\frac{N}{N_F-N}$ & 
$\frac{N}{N_F}$ \\
$M^i_j$ & $-$ & $N_F$ & $\overline{N}_F$ & $0$ & $2 \frac{N_F - N}{N_F}$\\
\hline
\end{tabular}
\end{center}
The  motivation behind the flavor transformation properties of 
the fields in the dual theory is that the gauge invariant 
polynomials

\begin{eqnarray} \label{DB}
B_{i_1 \cdots i_{N_D}}  &=&
    q_{i_1}^{\a_1} q_{i_2}^{\a_2} \cdots q_{i_N}^{\a_{N_D}} 
\epsilon_{\a_1 \a_2
\cdots \a_{N_D}} \\ \label{DBt} 
\tilde{B}^{j_1 \cdots j_{N_D}} &=& \tilde{q}_{\a_1}^{j_1}
 \tilde{q}_{\a_2}^{j_2}
 \cdots
\tilde{q}_{\a_{N_D}}^{j_{N_D}} 
\epsilon^{\a_1 \a_2 \cdots \a_{N_D}},
\end{eqnarray}
can be identified with the fields~(\ref{B}, \ref{Bt}) of the original 
 theory. The identification of~(\ref{M}) 
with the gauge singlets $M^i_j$ of the dual theory completes 
the isomorphism $I: V_D \to V$. The gauge invariants  
$q_i^{\a} \tilde{q}^j_{\a} $ are trivial on $U^{W_D}$, 
as $W_D = M^i_j q_i^{\a} \tilde{q}^j_{\a} $. 
We can show flavor anomaly matching for these theories 
using only two points, $\phi_1$ and $\phi_2$, in 
the above argument. $\phi_1$ has  coordinates 
\begin{equation} \label{p}
Q^{i\alpha} = \left\{ \begin{array}{ll}
m \delta^{i\alpha} & i \le N\\
0 & i > N
\end{array}\right. , \qquad
\tilde Q_{j \alpha} = 0.
\end{equation}
Note that $G\phi_1 = \{ (Q^{i \a},\tilde{Q}_{\b j})\; |\;
 \text{det}_{i \leq \a} 
Q = m^N, \tilde{Q}_{\b j} = Q^{i \a} = 0, i>\a \}$ is a closed set, 
and that $G$ is totally broken at $\phi_1$, 
then theorems I and II  apply at $\phi_1$.
The point $\pi(\phi_1) = \hat \phi_1$ in the IR theory is described by gauge
invariant meson and baryon  fields,
\begin{equation}\label{phiM}
M^i_j =0, \qquad \tilde B^{j_1\cdots j_{N_D}} =0,
\qquad B_{i_1 \cdots i_{N_D}} =
m^{N} \ \epsilon_{12...N_c i_1 \cdots i_{N_D}}.
\end{equation}
The unbroken flavor group at $\hat \phi_1$ is $F_1 = SU(N)_L \times
SU(N_D)_L \times SU(N_F)_R \times U(1)_B' \times U(1)_R'$, 
where (the Lie algebra of) 
$U(1)_B' \times R'$ is a linear combination 
of the original baryon and $R$ symmetry generators and generators of 
$SU(N_F)_L$. Under these
unbroken symmetries, the fields transform as

\begin{center}
\begin{tabular}{|c|c|c|c|c|c|}
\hline
 & $SU(N)_L$ & $SU(N_D)_L$ & $SU(N_F)_R$ & $U(1)_B'$ & $U(1)_R'$ \\
\hline
$Q^{i\alpha}$, $i \le N$ & $N$ & $-$ & $-$ & $0$ & $0$ \\
$Q^{i \alpha}$, $i > N$ & $-$ & $N_D$ & $-$ & $- N_F$ &
$(2N_F-2N)/(2N_F-N)$ \\
$\tilde Q_{j \alpha}$ & $-$ & $-$ & $\overline{N}_F$ & $N_D$ &
$(2N_F-2N)/(2N_F-N)$ \\
\hline
\end{tabular}
\end{center}

A natural choice for $\phi_1^D$ in the dual theory is
\begin{equation} \label{pd} 
\label{choicedual}
q_i^{\a} = \left\{ \begin{array}{ll}
m \delta^{\a}_{i-N} & i >  N \\
0 & i \leq  N
\end{array}\right. , \qquad
\tilde q_{\a}^j = 0, \qquad M^i_j = 0 
\end{equation}
Although the classical moduli spaces 
of these theories are different ($M^i_j$ can be arbitrary 
in the dual theory, whereas rank $M^i_j \leq N$ in the 
original theory), we can use theorem~I to check 
that, thanks to the superpotential $W_D$ in the dual 
theory,  the tangents at the shared point (\ref{phiM}) agree, 
as $\pi'_{\phi_1}(U) = \pi'_{\phi_1^D}(T_{\phi_1^D}U_D^W).$ 
This is the span of  $\delta M^i_j, i \leq N$ and the  $\delta 
B_{i_1,...,i_{N_D}}$ 
with at most one of the $i_k$ less than or equal to $N$.
Note that 
 $T_{\phi_1^D}U_D^W = \text{ker}\;  (\partial W / (\partial (\phi^D)^i
\partial (\phi^D)^j))\; |_{\phi^D_1} = \text{span} (\delta M^i_j (i \leq N), 
\delta q^{\alpha}_i)$, and that ker~$\pi'_{\phi_1^D}$ is the 
subspace   
$T_{\phi_1^D}U_D^W \cap \text{ker} \; (\partial \hat \phi ^i / 
\partial (\phi^D)^j)) \mid_{\phi_1^D} = \lie G \phi_1^D$, as anticipated 
by Theorem~I. \footnote{When $W \neq 0$ it may be  easier to check 
the equality ker $\pi'_{\phi_0} = \lie G \phi_0$ by 
using ker~$\pi'_{\phi_0} = T_{\phi_0}U^W \cap \text{ker} \;  
(\partial \hat{\phi} ^i / \partial \phi^j) \mid_{\phi_0}$,  
rather than proving that $U^W$ is irreducible and $\phi_0$ smooth 
 to apply  theorem~I.} 
The point $\phi_2$ is taken to be  ``symmetric'' 
to $\phi_1$ (i.e., with the roles of $Q$ and 
$\tilde{Q}$ exchanged). 
The matching of the full flavor group anomalies then 
follows from the matching of $F_1$ and $F_2$ anomalies~\cite{dm2}.\\

\noindent
{\bf Anomaly matching in dual $SU(N)$ theories with 
adjoint matter:} As in the QCD case, not every point 
in $\mc$ describes correctly the massless particle spectrum 
in the IR, 
but only those predicted by theorem~II. A dual theory 
based on the gauge group $SU(N_D'), N_D' = 2N_F -N$, 
$N_F $ flavors of fundamentals and conjugate fields, 
an adjoint and additional singlets is believed to describe 
the vacuum at the origin, where the full flavor symmetry 
group $SU(N_F)_L \times SU(N_F)_R \times U(1)_B \times R$ 
is unbroken. 
The field content and  transformation 
properties for both theories are  summarized in the table below~\cite{adjoint}

\noindent
\begin{center}
\begin{tabular}{|c|c|c|c|c|c|}
\hline
 & $G_{(D)}$ & $SU(N_F)_L$ & $SU(N_F)_R$ & $U(1)_B$ & $R$ \\
\hline
$Q^{i\alpha}$ &$N$ &$N_F$ & $-$ & $1$ & $1 - \frac{2N}{3N_F}$ \\
$\tilde Q_{j \alpha}$ & $\overline{N}$ & $-$ & $\overline{N}_F$ & $-1$ &
$1 - \frac{2N}{3N_F}$ \\ 
$X$ & Adj & $-$ & $-$  & $0$ & $\frac{2}{3}$\\ 
$T$ & $-$ & $-$ & $-$ & $0$ & $\frac{4}{3}$ \\ \hline 
$q^{\a}_i$ & $N_D'$ & $\overline{N}_F$ & $-$ & $\frac{N}{N_D'}$ & 
$1 - \frac{2N_D'}{3N_F}$\\
$\tilde{q}^j_{\b}$ & $\overline{N}_D'$ & $-$ & $N_F$ & $-\frac{N}{N_D'}$ & 
$1 - \frac{2N_D'}{3N_F}$ \\
$Y$ & Adj & $-$ & $-$ & $0$ & $\frac{2}{3}$\\
$M^i_j$ & $-$ & $N_F$ & $\overline{N}_F$ & $0$ & $2 -\frac{4N}{3N_F}$\\
$N^i_j$ & $-$ & $N_F$ & $\overline{N}_F$ & $0$ & $\frac{8N_F - 4N}{3N_F}$\\
$T$ & $-$ & $-$ & $-$ & $0$ & $\frac{4}{3}$ \\ \hline
\end{tabular}
\end{center}
We can identify the invariant fields [\ref{BB},\ref{BBt}] with 
\begin{eqnarray} \label{BBd}
B^{i_1,...,i_{n_1};j_1,...,j_{n_2}} &=& \frac{1}{N!} \epsilon^{i_1,...,
i_{n_1},l_1,...,l_{n^D_2}} \epsilon^{j_1,...,j_{n_2}, k_1,...,k_{n^D_1}} 
\nonumber \\ 
&& q_{k_1}^{\a_1} q_{k_2}^{\a_2} \cdots q_{k_{n_1^D}}^{\a_{n_1^D}} 
Y^{\b_1}_{\g_1} \cdots Y^{\b_{n^D_2}}_{\g_{n_2^D}} q_{l_1}^{\g_1} 
\cdots q_{l_{n^D_2}}^{\g_{n_2^D}} \epsilon_{\a_1 \cdots \a_{n_1^D} 
\b_1 \cdots \b_{n_2^D}}, \\ \label{BBtd}  
\tilde{B}_{i_1,...,i_{n_1};j_1,...,j_{n_2}} &=& \frac{1}{N!} \epsilon_{i_1,...
, i_{n_1}; l_1,...,l_{n^D_2}}  \epsilon_{j_1,...,j_{n_2};k_1,...,k_{n_1^D}} 
\nonumber \\
&& \tilde{q}_{\a_1}^{k_1} \tilde{q}_{\a_2}^{k_2} \cdots 
\tilde{q}_{\a_{n^D_1}}^{k_{n^D_1}} 
Y_{\b_1}^{\g_1} \cdots Y_{\b_{n_2^D}}^{\g_{n_2^D}} \tilde{q}_{\g_1}^{l_1} 
\cdots \tilde{q}_{\g_{n^D_2}}^{j_{n^D_2}} \epsilon^{\a_1 \cdots \a_{n_1^D} 
\b_1 \cdots \b_{n_2^D}}, 
\end{eqnarray}
where $n^D_1 = N_F - n_2, \; n^D_2 = N_F - n_1$. 
The identification of the invariants 
(\ref{Ma}), (\ref{N}) and (\ref{T}) with the singlets 
$M^i_j$, $N^i_j$ and $T$ of the 
dual theory completes the map $I: V_D \to V$. A superpotential
\begin{equation} \label{W}
W_D = \frac{1}{3} \text{tr} Y^3 + M^i_j \tilde{q}^j_{\a} q_i^{\a} 
+ N^i_j \tilde{q}^j_{\a} Y_{\b}^{\a} q^{\b}_i - \frac{T}{N} \tr Y^2 
\end{equation}
is added to the dual theory, then the gauge invariants 
$\tilde{q}^j_{\a} q_i^{\a}$ and  
$\tilde{q}^i_{\a} Y_{\b}^{\a} q^{\b}_i$ need not be considered, 
as they are trivial on $U_D^{W_D}$. Also, $\tr Y^2 = T$ on $U^{W_D}_D$.  
Note that $N + s = N_F = N_D - s$. As duality is an involutive 
operation we can restrict ourselves to the study of the $s \leq 0$ case.
To understand anomaly matching we can restrict  further to the 
self dual case 
$s = 0$, as we can flow to the other cases by adding a 
mass term to the ``electric'' theory to decouple 
a flavor. Duality is compatible with this flow, which 
also preserves anomaly matching~\cite{dm,dm2}. The motivation  
behind considering the $s = 0 $ case is that more similarities 
between the classical chiral rings of the original and dual 
theories are to be expected in this case~\cite{adjoint}. 
Finally,  assume for simplicity $N = 2n$ (Higgs effect allows 
us to flow to the odd $N$ case) and 
consider the point $\phi_1$ of coordinates 
\begin{equation} \label{pa}
X = m \left(\begin{array}{cc}
    0 & 1 \\
    0 & 0 \end{array} \right), \hspace{1cm}
Q = m \left(\begin{array}{cc}
    0 & \sqrt{2} \\
    0 & 0 \end{array} \right), \hspace{1cm} \tilde{Q} =0.
\end{equation}
In the above  matrix notation upper or left indices 
label rows,  and $X,Q$ and $\tilde{Q}$ 
are broken up in square matrices of size $n$. 
$\phi_1$ is a smooth point in the irreducible 
set $U^W$ which breaks $G = SL(n,\c)$ completely and is $D$ flat, 
then theorems I and II apply (see however footnote 3). 
The only nonzero coordinates of $\hat \phi_1 = 
\pi(\phi_1)$ are  $B^{i_1,...,i_n;j_1,...,j_n} = 2^n m^{3n} \epsilon^{i_1,...,
i_n,n+1,...,2n} \epsilon^{j_1,...,j_n,n+1,...,2n}$. 
We choose $\phi_1^D$ to be the point with coordinates 
\begin{equation} \label{pad}
Y = \a m \left(\begin{array}{cc}
    0 & 0 \\
    1 & 0 \end{array} \right), \; \; 
q = \a m \left(\begin{array}{cc}
    0 & \sqrt{2} \\
    0 & 0 \end{array} \right), \; \; \tilde{q} = M = N = 0, 
\end{equation}
where $\a ^{3n} = -N!/(n!)^4$. 
It is straightforward to verify that $\phi _1^D$ is in $U_D^{W_D}$ 
and satisfies the hypothesis of the matching theorem. 
The tangents $T_{\hat \phi_1} \mc$ and $T_{\hat \phi_1} \mc ^D$ 
both equal the span of $\delta M^i_j, i \leq n,$ 
and $\delta N^i_j, i \leq n$, $\delta T$, 
and the fields $\delta B^{i_1,...,i_n,j_1,...,j_n}$ 
with at most one index bigger than $n$ (the nonzero fields 
$\delta B^{i_1,...,i_{n_1},j_1,...,j_{n_2}}$  with 
$n_1 = n \pm 1$ and at most one index bigger than $n$ 
are linearly dependent from these). 
The vacuum $\hat \phi_1$ breaks $F$ to $F_1 = SU(n)_L \times SU(n)_L \times 
SU(N)_R \times 
U(1)_B' \times R'$, with $U(1)_B' \times R'$ a combination
 of $U(1)_B \times R$ and $SU(N)_L$. We can take 
flavor rotated versions of~(\ref{pa},\ref{pad}), or just the 
``symmetric'' point with $Q^{i \a} = 0$ to complete the proof 
of anomaly matching.\\

Our  results allow the prediction of the dimension 
of the dual group $G_D$. Assume the matter content and superpotential of 
the dual theory are obtained, e.g, from the rules in  \cite{es}), 
up to the value of $N_D$. Our anomaly matching mechanism implies 
\begin{equation} \label{dim}
\text{dim} \; \mc = \text{dim} \;  \mc ^D. 
\end{equation}
Any two (unrelated) points 
$\phi_0 \in U^W$ and $\phi_0^D \in U_D^{W_D}$ 
satisfying the hypothesis of Theorem~I~(d) can be used to calculate 
these dimensions
\begin{equation} \label{dim2}
\text{dim} \; \mc = \text{rank} \; \pi'_{\phi_0} = \text{dim} \; T_{\phi_0}U^W 
- \text{dim} \; G = \text{dim} \; U^W - \text{dim} \;  G, 
\end{equation}
and analogously for $\mc^D$.
This calculation can be done even before 
checking  any connection between $V$ and $V_D$. 
In the QCD example, using the points of Eqs.~(\ref{p}, \ref{pd}), 
we obtain dim $U = 2 N N_F$, dim $U^W =$ dim $T_{\phi_0}U^W$. 
The latter is spanned by the $N_D N_F$ $\delta q^{\a}_i$ 
and the $(N_F - N_D) N_F$ $\delta M^i_j, i \leq N_F - N_D,$
then~(\ref{dim},\ref{dim2}) give
 the following equation on the indeterminate $N_D$:
$$2 N N_F - (N^2 -1) = N_D N_F + (N_F - N_D) N_F - (N_D^2 -1).$$
The solutions of this equation are $N_D = \pm (N_F -N)$. For the theory 
with matter in the adjoint, we analyze the case $N$ even and 
assume $N_D$ is also even, with $N_F > N/2,N_D/2$. To calculate 
dimensions we use suitable generalizations 
of Eqs.~(\ref{pa},\ref{pad}) to the $N_F \neq N$ case, 
where now the lower $Q^{i \a}$ blocks are $(N_f -N/2) \times N/2$ 
matrices and the left hand side $q^{\a}_j$ blocks 
are $(N_D/2) \times (N_F - N_D/2)$ matrices. The tangent 
$T_{\phi_0}U^W$ is spanned by $N^2/2$ fields $\delta X^{\a}_{\b}, \;
\delta T,$ and the unconstrained $\delta Q^{i \a}$ fields, so we get
\begin{equation}\label{dma}
\text{dim} \; \mc  =  \left(\frac{N^2}{2}\right) + 1 
+ 2N N_F - (N^2 -1).
\end{equation}
The tangent $T_{\phi_0^D} U_D^{W_D}$ is spanned by 
 $N_D^2/2$ fields $\delta Y^{\a}_{\b}, \;
\delta T,$ the unconstrained $\delta Q^{i \a}$, and the 
$\delta M^i_j$ and $\delta N^i_j$ with $i \leq N_F - N_D/2$, 
therefore 
\begin{equation} \label{dmad} 
 \text{dim} \; \mc^D  = \left(\frac{N_D^2}{2}\right) + 1
+ N_D N_F  + 2 \left( N_F - \frac{N_D}{2} \right) N_F - (N_D^2 -1).
\end{equation}
Equating (\ref{dma}) and (\ref{dmad}) we obtain $N_D = \pm (2N_F -N)$, 
as expected.\\

Understanding flavor anomaly matching for other pairs 
of dual theories like those involving $SO(N)$ and $SP(2N)$ 
gauge groups requires stronger versions of theorem II 
which are currently under study, 
their treatment seems to be analogous to the 
simpler cases presented here. 
The studied examples suggest that the satisfaction of `t~Hooft's 
consistency conditions is to be expected from the 
similarities of the classical moduli spaces of dual 
theories, and does not constitute an independent test on 
the duality hypothesis.\\

I would like to thank K.~Intriligator, A.~Manohar,  
E.~Poppitz and W.~Skiba
for useful discussions and comments on the manuscript. 
This work was supported in part by a Department of Energy grant 
DOE-FG03-97ER40546.

\end{document}